# Strong-coupling expansion of lattice O($N$) $\sigma$ models

Massimo Campostrini, Andrea Pelissetto, Paolo Rossi, and Ettore Vicari,
Dipartimento di Fisica dell'Università and I.N.F.N., I-56126 Pisa, Italy

We report progress in the computation and analysis of strong-coupling series of two- and three-dimensional O($N$) $\sigma$ models. We show that, through a combination of long strong-coupling series and judicious choice of observables, one can compute continuum quantities reliably and with a precision at least comparable with the best available Monte Carlo data.

## 1. Computation of strong-coupling series

We perform the computation of strong-coupling series applying the method presented in Ref. [1]. It involves essentially the calculation of a *geometrical factor*, common to all spin models admitting a character-like expansion, and of a *group-theoretical factor*, independent of the lattice connectivity.

We focused on the two-point fundamental Green's functions $G(x, y)$ and on the free energy $F$. Their geometrical factors are computed by generating all paths connecting $x$ and $y$ in the first case and all closed paths in the second case, and then reducing each path to a group-theoretical diagram; this is done by an ad-hoc computer program. The group-integration technique for O($N$) models is covered in Ref. [2].

The orders of the expansion reached so far are presented in Table 1. In order to generate longer series for the wall-wall correlation length $\xi_w$, we computed higher orders of selected large-distance Green's functions. Computation is in progress for the two-point Green's functions in the adjoint representation. So far we computed it on the square lattice to 20th order.

To give an idea of the complexity of this computations, the evaluation of all the two-point fundamental Green's functiond to 15th order involves $\simeq 24 \times 10^9$ nontrivial paths.

In order to check and compare with our strong-coupling results, we computed the large-$N$ limit of the nearest-neighbour O($N$) $\sigma$ models on the relevant lattices and verified that its strong-coupling expansion concides with the large-$N$ limit of our strong-coupling series.

Table 1
Order of computed strong-coupling series

| lattice | $G$ | $F$ | $\xi_w$ |
|---|---|---|---|
| cubic | 15 | 16 | 12 |
| diamond | 21 | 24 | 13 |
| square | 21 | 24 | 16 |
| honeycomb | 30 | 32 | 25 |

## 2. Physical quantities

From the Green's function we can derive many interesting quantities. On each lattice we compute

$$m_{2j} = \sum_x \left(x^2\right)^j G(x), \qquad \chi = m_0,$$

$$\xi_G = \frac{m_2}{2d\chi}, \qquad M_G = \frac{1}{\xi_G}, \qquad \omega = \frac{m_2^2}{\chi m_4} \quad (1)$$

Furthermore, on the square lattice we define

$$M_s^2 = 2(\cosh \mu_s - 1),$$
$$\text{where} \quad \tilde{G}^{-1}(p_1 = i\mu_s, \, p_2 = 0) = 0, \quad (2)$$

$$M_d^2 = 4\left(\cosh \frac{\mu_d}{\sqrt{2}} - 1\right),$$
$$\text{where} \quad \tilde{G}^{-1}\left(p_1 = i\frac{\mu_d}{\sqrt{2}}, \, p_2 = i\frac{\mu_d}{\sqrt{2}}\right) = 0. \quad (3)$$

$\mu_s$ and $\mu_d$ determine the long-distance exponential decay of $G(x)$ on the side and on the diagonal respectively. In the continuum limit $M_s = M_d = 1/\xi$.

Similar quantities are defined for the other lattices. In particular, on the honeycomb lattice we define $M_h^2$, the equivalent of $M_s^2$.

In a generalized lattice Gaussian model, and therefore in the large-$N$ limit of O($N$) $\sigma$ models, the following results hold:

$$\frac{M_s^2(\beta)}{M_G^2(\beta)} = \frac{M_h^2(\beta)}{M_G^2(\beta)} = 1,$$

$$\omega(\beta) = \left(4 + \frac{M_G^2(\beta)}{4}\right)^{-1}. \quad (4)$$

We can transform a series in powers of $\beta$ in a series in powers of the energy $E$, since we computed to high orders $E(\beta) = \beta + O(\beta^3)$.

## 3. Strong-coupling analysis of the 2-$d$ O(3) $\sigma$ model.

Details of the strong-coupling analysis we describe in this section can be found in Ref. [3].

In asymptotically free models, where $\beta_c = \infty$, the task of determining physical continuum quantities from a strong-coupling approach appears difficult. Nevertheless, as we shall see below, the strong-coupling analysis provides quite accurate continuum limit estimates when applied directly to dimensionless renormalization-group invariant ratios of physical quantities, essentially by exploiting the following ideas:

(i) Let us indicate with $R(\beta)$ a dimensionless renormalization-group invariant quantity. Since at sufficiently large $\beta$ $R(\beta)$ behaves as

$$R(\beta) - R^* \sim \frac{1}{\xi^2}, \quad (5)$$

where $R^*$ is the fixed point and therefore continuum value, a reasonable estimate of $R^*$ may be obtained at $\beta$-values corresponding to large (but finite) correlation lengths, where the curve $R(\beta)$ should be already stable (scaling region). This is the same idea underlying Monte Carlo studies.

(ii) Another interesting possibility is to change variable from $\beta$ to the energy $E$, and analyze the series in powers of $E$, which are obtained by inverting the strong-coupling series of the energy $E = \beta + O(\beta^3)$ and substituting into the original series in powers of $\beta$. It should be easier to reach the continuum limit this way, since it occurs at a finite value of $E$, i.e. for $E \to 1$.

Furthermore dimensionless renormalization-group invariant ratios of physical quantities are expected to have a simpler analytical structure (in the $\beta$ or $E$ complex plane), which may be better approximated by standard Padè-type approximants.

The analysis of both the strong-coupling series calculated on the square and honeycomb lattices offers us the possibility of testing universality, which represents also a further check for possible systematic errors in the analysis employed.

In our analysis of the strong-coupling series we constructed $[l/m]$ Padé approximants (PA's) and Dlog-PA's of both the series in $\beta$ and in the energy. While simple $[l/m]$ PA's provide directly the quantity at hand, in a Dlog-PA analysis one gets a $[l/m]$ approximant by reconstructing the original quantity from the $[l/m]$ PA of its logarithmic derivative. Continuum estimates are then obtained by evaluating the approximants of the energy series at $E = 1$, and those of the $\beta$ series at a value of $\beta$ corresponding to a reasonably large correlation length.

As final estimates we take the average of the results from the non-defective PA's using all available terms of the series. The errors we display are just indicative, they are the variance around the estimate of the results coming from PA's using also a few less terms of the series, which should give an idea of the spread of the results coming from different PA's. Such errors do not always provide a reliable estimate of the systematic errors, which may be underestimated especially when the structure of the function (or of its logarithmic derivative) is not well approximated by a meromorphic analytic function. In such cases a more reliable estimate of the systematic error would come from the comparison of results from the analysis of different series representing the same quantity, which in general are not expected to have the same structure.

By rotation invariance the ratio $r \equiv M_s^2/M_d^2$ (on the square lattice) should go to one in the continuum limit. Therefore the analysis of such ratio should be considered as a test of the procedure employed to estimate continuum physical quantities. From $G(x)$ up to $O(\beta^{21})$ we could calculate the ratio $r$ up to $O(\beta^{14})$. Our final estimates for $r^*$, the value of $r$ at the continuum limit, are $r^* = 0.9997(13)$ from the energy analysis, and $r^* = 1.0001(6)$ from the $\beta$ analysis performed at $\beta = 0.55$, corresponding to a correlation length $\xi \simeq 25$. The precision of these results is remarkable.

Calculating a few $G(x)$ up to 23rd order, we obtained the ratio $s \equiv M_s^2/M_G^2$ up to $16^{\text{th}}$ order on the square lattice. No exact results are known about the continuum limit $s^*$ of the ratio $s$, except for its large-$N$ limit: $s^* = 1$. Both large-$N$ and Monte Carlo estimates indicate a value very close to one. From a $1/N$ expansion [4]:

$$s^* = 1 - \frac{0.006450}{N} + O\left(\frac{1}{N^2}\right). \qquad (6)$$

Monte Carlo simulations at $N = 3$ [5] gave the estimate $s^* = 0.9985(12)$.

The analysis of the strong-coupling series of $s$ leads to $s^* = 0.999(3)$ from the $E$-approximants, and $s^* = 0.998(1)$ from the $\beta$ approximants evaluated at $\beta = 0.55$, in full agreement with the estimates from the $1/N$ expansion and Monte Carlo simulations. With increasing $N$, the central estimate of $s^*$ gets closer to 1.

On the honeycomb lattice, $G(x)$ up to $30^{\text{th}}$ order allows to calculate $s_h \equiv M_h^2/M_G^2$ up to $20^{\text{th}}$ order. The analysis of the energy series yields the estimate $s_h^* = 0.997(3)$, in agreement with the result from the square lattice.

On the square lattice, the analysis of the strong-coupling series of $\omega \equiv m_2^2/(\chi m_4)$ leads to the estimates $\omega^* = 0.2498(6)$ from the energy analysis at $E = 1$, and $\omega^* = 0.2499(6)$ from the $\beta$ analysis at $\beta = 0.55$. On the honeycomb lattice we estimated $\omega^* = 0.248(3)$. Again, universality is confirmed.

The comparison with the exact $N = \infty$ calculations shows that quantities like $s$ and $\omega$, which describe the small momentum universal behaviour of $\widetilde{G}(p)$ in the continuum limit, change very little from $N = 3$ to $N = \infty$, indicating that the two point function is substantially Gaussian at small momentum. Differences must eventually appear at large momentum, as predicted by simple weak coupling calculations supplemented by a renormalization group resummation. But the large momentum regime is hardly reachable by a strong-coupling analysis. Important differences are however present in other Green's functions even at small momentum, as shown in the analysis of the four-point zero-momentum renormalized coupling, whose definition involves the zero-momentum four-point correlation function [6].

## 4. Strong-coupling analysis of the 2-$d$ XY model.

The two dimensional $XY$ model is conjectured to experience a Kosterlitz-Thouless critical phenomenon, characterized by an exponentially divergent correlation length. Setting $\tau \equiv \beta_c - \beta$,

$$\xi \sim \exp\left(\frac{b_\xi}{\tau^\sigma}\right), \qquad (7)$$

$$\chi \sim \tau^\theta \exp\left(\frac{b_\chi}{\tau^\sigma}\right) \sim \xi^{2-\eta}. \qquad (8)$$

A renormalization group analysis applied to the Coulomb gas model predicts: $\eta = \frac{1}{4}$, $\sigma = \frac{1}{2}$ and $\theta = \frac{1}{16}$.

Support to the presence of this phenomenon has been provided by Monte Carlo techniques (cfr. e.g. Refs. [8–10]), and by strong-coupling expansion method ((cfr. e.g. Ref. [11]). Since we computed the series on the honeycomb lattice and extended the series on the square lattice series up to $O(\beta^{21})$, we performed a new strong-coupling analysis.

In order to check the Kosterlitz-Thouless critical behaviour we have analyzed the strong-coupling series of $\ln \chi$ and $\ln(\xi_G^2/\beta)$, which should behave as

$$\ln \chi \sim \ln(\xi_G^2/\beta) \sim \tau^{-\sigma}. \qquad (9)$$

A zero value for the exponent $\sigma$ would indicate a standard power-law critical behaviour. Beside PAs, we also employed integral approximants (IA's) [7], which allow a more general analysis as they can reproduce a larger class of behaviours close to criticality, reducing possible systematic errors in the resummation of the series. On the other hand, in order to get stable and therefore acceptable results, IA's require in general more terms in the series to be resummed than Dlog-PA's. We have observed this fact also in our strong-coupling analysis; indeed often Dlog-PA's turned out to be more stable than IA's, but subject to a larger systematic error, which emerged from the comparison of results from different series.

From an unbiased IA analysis of the series of $\ln \chi$ for the square lattice we found $\beta_c = 0.558(2)$ and $\sigma = 0.5(1)$, which strongly support the KT critical behaviour. The same analysis on the honeycomb lattice led to $\beta_c = 0.877(3)$ and $\sigma = 0.4(1)$.

We also performed biased IA analysis determining the value of $\beta_c$ such that $\sigma = \frac{1}{2}$, leading to a biased estimate: $\beta_c = 0.559(1)$, which is in agreement with a corresponding biased analysis of Monte Carlo data: $\beta_c = 0.559(3)$ [8], and with a quite precise Monte Carlo renormalization group determination of $\beta_c$ [10]: $\beta_c = 0.5599(3)$. On the honeycomb lattice, by the same analysis, we found $\beta_c = 0.881(2)$.

In order to determine the exponent $\eta$ we considered the quantity

$$A_\eta \equiv 2\left(1 - \frac{\ln \chi}{\ln(\xi_G^2/\beta)}\right) \simeq \eta + O(\tau^\sigma) \quad (10)$$

close to $\beta_c$. Resumming the corresponding series by PA's and Dlog-PA's and evaluating it at $\beta_c \simeq 0.559$, one finds a quite stable result: $\eta = 0.228(1)$. This result is confirmed on the honeycomb lattice: $\eta = 0.231(2)$. These estimates of $\eta$ do not agree with the expected value $\eta = \frac{1}{4}$, they are about 10% far from it. On the other hand when analyzing the series in the energy we got again a rather stable result but $\eta = 0.207(5)$, indicating the presence of a sizeable systematic error of about 10%. The agreement between the square and honeycomb lattice results may be accidental, and may be partially explained by the fact that the origin of the systematic error in the analysis should be similar.

A source of systematic error may be the $O(\tau^\sigma)$ correction expected in Eq. (10) which cannot be reproduced by PA's or Dlog-PA's. Eq. (10) implies a behaviour

$$\text{Dlog}\, A_\eta(\beta) \sim (\beta_c - \beta)^{\sigma-1} \quad (11)$$

close to $\beta_c$. In the Dlog-PA's the above singularity should be mimicked by a shifted pole at a $\beta$ larger than $\beta_c$. Indeed in the analysis of the series of $A_\eta$ we have found a singularity typically at $\beta \simeq 1.1 \div 1.2\ \beta_c$. This fact will eventually affect the determination of $A_\eta$ close to $\beta_c$ by a systematic error. However since the singularity is integrable the error must be finite, and the analysis shows that such errors are actually reasonably small. The behaviour (10) could be reproduced by IA's, but we did not get sufficiently stable and thus acceptable results from them.

In conclusion the strong-coupling analysis of order strong-coupling expansion of $G(x)$ on the square and honeycomb lattices substantially supports the Kosterlitz-Thouless critical phenomenon.

## REFERENCES

1. M. Campostrini, P. Rossi, and E. Vicari, Phys. Rev. D52 (1995) 358.
2. M. Campostrini et al., Application of the $O(N)$ hypersferical harmonics..., these proceedings.
3. M. Campostrini, A. Pelissetto, P. Rossi, and E. Vicari, Strong-coupling analysis of the $O(N)$ $\sigma$ models with $N \geq 3$, Pisa preprint IFUP-TH 47/95.
4. P. Biscari, M. Campostrini and P. Rossi, Phys. Lett. 242B (1990) 225; H. Flyvbjerg, Nucl. Phys. B348 (1991) 714.
5. S. Meyer, unpublished.
6. M. Campostrini, A. Pelissetto, P. Rossi and E. Vicari, Four-point renormalized coupling constant in $O(N)$ models, Pisa preprint IFUP-TH 24/95, hep-lat 9506002.
7. A. J. Guttmann and G. S. Joyce, J. Phys. A5 (1972) L81; D. L. Hunter and G. A. Baker Jr., Phys. Rev. B49 (1979) 3808.
8. R. Gupta, and C. F. Baillie, Phys. Rev. D45 (1992) 2883.
9. L. Biferale and R. Petronzio, Nucl. Phys. B328 (1989) 677.
10. M. Hasenbush, M. Marcu, and K. Pinn, Physica A208 (1994) 124.
11. P. Butera and M. Comi, Phys. Rev. B47 (1993) 11969.